\documentclass[draft]{agujournal2019}
\usepackage{url} 
\usepackage{lineno}
\usepackage{soul}
\usepackage{rotating}
\draftfalse

\journalname{Geophysical Research Letters}
\begin{document}
	
%
%

\title{Field-Aligned Currents and Auroral Precipitation During the Terrestrial Alfv\'{e}n Wing State}

%
%



\authors{B. L. Burkholder\affil{1,2}, L.-J. Chen\affil{2}, D. Lin\affil{3}, S. K. Vines\affil{4},  K. A. Sorathia\affil{5}, C. F. Bowers\affil{6}}

\affiliation{1}{Goddard Planetary Heliophysics Insitute, University of Maryland Baltimore County, Baltimore, MD, USA}
\affiliation{2}{Heliophysics Science Division, NASA Goddard Space Flight Center, Greenbelt, MD, USA}
\affiliation{3}{High Altitude Observatory, National Center for Atmospheric Research, Boulder, CO, USA}
\affiliation{4}{Southwest Research Institute, San Antonio, TX, USA}
\affiliation{5}{The Johns Hopkins University Applied Physics Laboratory, Laurel, MD, USA}
\affiliation{6}{School of Cosmic Physics, DIAS Dunsink Observatory, Dublin Institute for Advanced Studies, Dublin, Ireland}





\correspondingauthor{Brandon Burkholder}{burkbran@umbc.edu}



\begin{keypoints}
	\item During the April 2023 storm, northern polar field-aligned current (FAC) flowed at magnetic local time 3-13from Earth's Alfv\'{e}n wings (AWs).
	\item Under B$_y$ dominant IMF, north hemisphere AW aurora was likely confined to the dawn day-side, but stretched to dusk before the wings formed.
	\item Simulated AW FACs are driven by localized flow shear at the wing edge, in a layer less than half as long as the pre-AW equivalent.
\end{keypoints}

%
%

%
%

\begin{abstract}
	When sub-Alfv\'{e}nic (Alfv\'{e}n Mach number M$_A<$1) plasmas impact Earth, Alfv\'{e}n wings (AWs) develop. A Multiscale Atmosphere Geospace Environment (MAGE) simulation of the April 2023 storm, validated by Active Magnetosphere and Planetary Electrodynamics Response Experiment (AMPERE) data, reveals the field-aligned-current (FAC) generation mechanism and predicts auroral precipitation for Earth's AWs. Simulation and observations show northern hemisphere planetward flowing electrons are predominantly at magnetic local times (MLTs) 8-13. Before the AWs formed, solar wind conditions were similar and M$_A\sim1.4$, yet the same FAC system extended from 9-18 MLT. Flow vorticity drives FACs at the boundary of the AWs and unshocked solar wind. The AW shape presents a different obstacle to the solar wind compared to typical lobe fluxes, producing the unique FAC distribution. New insights aboutAW FACs and precipitating electron energy flux will help understand auroral features for exoplanets inside their host star's Alfv\'{e}n zone.
\end{abstract}

\section*{Plain Language Summary}
The Earth is typically enveloped in a flow of plasma emanating from the sun that produces a supersonic boom-like structure (a bow shock) in front of Earth. In rare instances when the shock is not present, Earth's magnetosphere develops so-called Alfv\'{e}n wings. When the Alfv\'{e}n wings form, the magnetospheric cross-section interacting with the solar wind takes on a different shape compared to the typical magnetosphere. The Earth's aurora is a product of electric currents that flow out of the upper atmosphere, and they are generated where the magnetosphere interacts with the solar wind and interplanetary magnetic field. Thus, the formation of Alfv\'{e}n wings during the April 2023 geomagnetic storm presents a unique opportunity to examine global electric current generation in an extreme regime, and better understand the structure and evolution of planetary and exoplanetary magnetospheres. By comparing simulations and observations of Earth's Alfv\'{e}n wings, the Alfv\'{e}n wing aurora is predicted to have occurred in a restricted sector of the day-side, where it could potentially have been observed with sensitive cameras, even though it may have been outshone by sunlight and not visible to human eyes.

%
%

%


\section{Introduction}
Global-scale, high-latitude field-aligned currents (FACs), also known as Birkeland currents \cite{Birkeland1908,Birkeland1913,Zmuda1966}, are generated in the magnetosphere and flow along magnetic field lines into the polar caps. They close via Hall and Pederson currents flowing in Earth's polar ionosphere, which are associated with atmospheric Joule heating \cite{Crowley1991}. FACs provide charge carriers to enhance ionospheric conductivity, including planetward flowing electrons which can be accelerated by double layers to produce aurora \cite{Andersson2012} (aurora can also be formed by wave particle interactions \cite{Ni2016} and Alfv\'{e}nic acceleration \cite{Chaston2004}). Earth's global current system carries MAs (millions of Amperes) during quiet times and 10s of MA during geomagnetic storms \cite{Pederson2021}. These produce magnetic perturbations on the ground and in space, which are hazardous to vital technologies if sufficiently intense \cite{Pilipenko2021}. 

The morphology and evolution of FACs is controlled by the solar wind-magnetosphere interaction \cite{Sato1979}. FACs connecting Earth's magnetosphere and ionosphere are classified into Region I and II current systems \cite{Iijima1976,Iijima1978}, mapping to currents originating on the magnetopause and in the ring current and/or magnetotail, respectively. Geomagnetic storms and substorms produce the strongest and most variable FACs (e.g., \citeA{Coxon2023}). Geomagnetic storms are associated with extreme variations of solar wind driving, such as the interplanetary coronal mass ejection (CME) shock impact preceding rarefied plasma in the magnetic cloud. Sustained intervals of sub-Alfv\'{e}nic (Alfv\'{e}n Mach number M$_A <$ 1, with M$_A$ the ratio of solar wind velocity, V$_{sw}$, to local Alfv\'{e}n speed, V$_A$) solar wind are often associated with rarefied CME magnetic clouds following fast and dense CME sheath material \cite{Lavraud2002,Hajra2022}. During sustained sub-Alfv\'{e}nic solar wind, the lack of bow shock exposes the magnetopause to unshocked solar wind \cite{Chen2024}, a faster flow than typical magnetosheath plasma. Along with the magnetospheric transformation to an Alfv\'{e}n wing (AW) configuration \cite{Ridley2007}, direct exposure of the magnetopause leads to unique behaviors of magnetic reconnection and FAC generation. 

Both day-side and night-side magnetic reconnection processes have been shown to be altered during the AW transformation \cite{Chen2024,Burkholder2024,Gurram2025}. This study examines the unique FAC distribution and predicted auroral signatures during the same event. By combining Multiscale Atmosphere Geospace Environment (MAGE) simulations and Active Magnetosphere and Planetary Electrodynamics Response Experiment (AMPERE) observations during the April 2023 storm, we demonstrate that upward FAC is localized in the dawn day-side sector. With precipitating electron energy flux as an auroral proxy, it can be predicted that the AW aurora also occurred predominantly in the same sector.

\section{MAGE and AMPERE}
The global geospace simulation is similar to \citeA{Sorathia2023} and \citeA{Sorathia2024}. The MAGE configuration includes: the global magnetohydrodynamic (MHD) model Grid Agnostic MHD for Extended Research Applications (GAMERA, \citeA{Zhang2019,Sorathia2020}); the Rice Convection Model (RCM; \citeA{Toffoletto2003}); and the RE-developed Magnetosphere-Ionosphere Coupler/Solver (REMIX; \citeA{Merkin2010}). The simulation uses Solar Magnetospheric (SM) coordinates (z-axis aligned to Earth's northern magnetic dipole and y-axis perpendicular to the Earth-Sun line towards dusk). GAMERA integrates the ideal MHD equations driven by time-dependent solar wind parameters from OMNIWeb (see Figures \ref{fig:fig1}a and b). The inner boundary condition is derived by REMIX from the distribution of the ionospheric electrostatic potential \cite{Merkin2010}. The simulation covers 00:00 April 23 - 23:58 April 24, whichincludes pre-storm preconditioning. The version of MAGE used here considers mono-energetic and diffuse electron precipitation \cite{Lin2021}, but does not capture the Alfv\'{e}nic aurora. Once electron precipitation fluxes are derived, Pedersen conductance is evaluated using the \citeA{Robinson1987} formula and Hall conductance is derived from Pedersen conductance using the empirical ratio from \citeA{Kaeppler2015}. During the transition from super-Alfv\'{e}nic to sub-Alfv\'{e}nic solar wind, the bow shock retreats upstream and eventually interacts with the boundary. To isolate the magnetosphere from reflected waves, the upstream boundary was moved to 120 R$_E$ \cite{Ridley2007,Chane2015,Wilder2019}. This is accomplished by stretching the grid, which reduces the resolution (compared to the typical GAMERA `Quad' grid) by a factor $\sim2$, resulting in the highest MHD grid resolution being $\sim$2400 km in the near-Earth plasma sheet (the same grid as \citeA{Burkholder2024}). MAGE is developed by the NASA DRIVE Science Center for Geospace Storms.

AMPERE \cite{Anderson2002,Anderson2014,Anderson2021} provides global estimates of high-latitude ionospheric radial current (j$_r$), which we use to validate MAGE polar FAC profiles. AMPERE data are provided in Altitude-Adjusted Corrected Geomagnetic (AACGM) coordinates, which reduce to SM coordinates to lowest order (accounting for about 95$\%$ of Earth’s magnetic field \cite{Laundal2017}). The native coordinate systems of MAGE and AMPERE are thus sufficiently similar for qualitative comparison. AMPERE j$_r$ has two minutes cadence, though 10-minutes accumulation is required for sufficient global coverage (i.e. revisit time). The AMPERE grid has 24 MLT sectors and extends 50$^\circ$ MLat from the pole (MAGE FAC output has 360 MLT sectors and extends 45$^\circ$ MLat from the pole). Because AMPERE j$_r$ is defined based on the radial direction, which is opposite the field-aligned direction in the northern hemisphere, a negative sign is applied to AMPERE data to facilitate comparison with MAGE. The small difference between $j_r$ and true FAC can be neglected here because the AW FAC structures are at high magnetic latitudes (MLat $>$60$^\circ$). At 780 km altitude, the field-aligned direction at 60$^\circ$ MLat is only 15$^\circ$ deviated from the radial direction (and less at higher MLat). We thus refer to AMPERE radial current as FAC (j$_{||}$) for simplicity. This study focuses on the northern hemisphere due to observational constraints of AMPERE in the southern hemisphere (see discussion in \citeA{Waters2020}). 

\section{April 2023 Geomagnetic Storm}
\begin{figure}
	\centering
	\includegraphics[width=\linewidth]{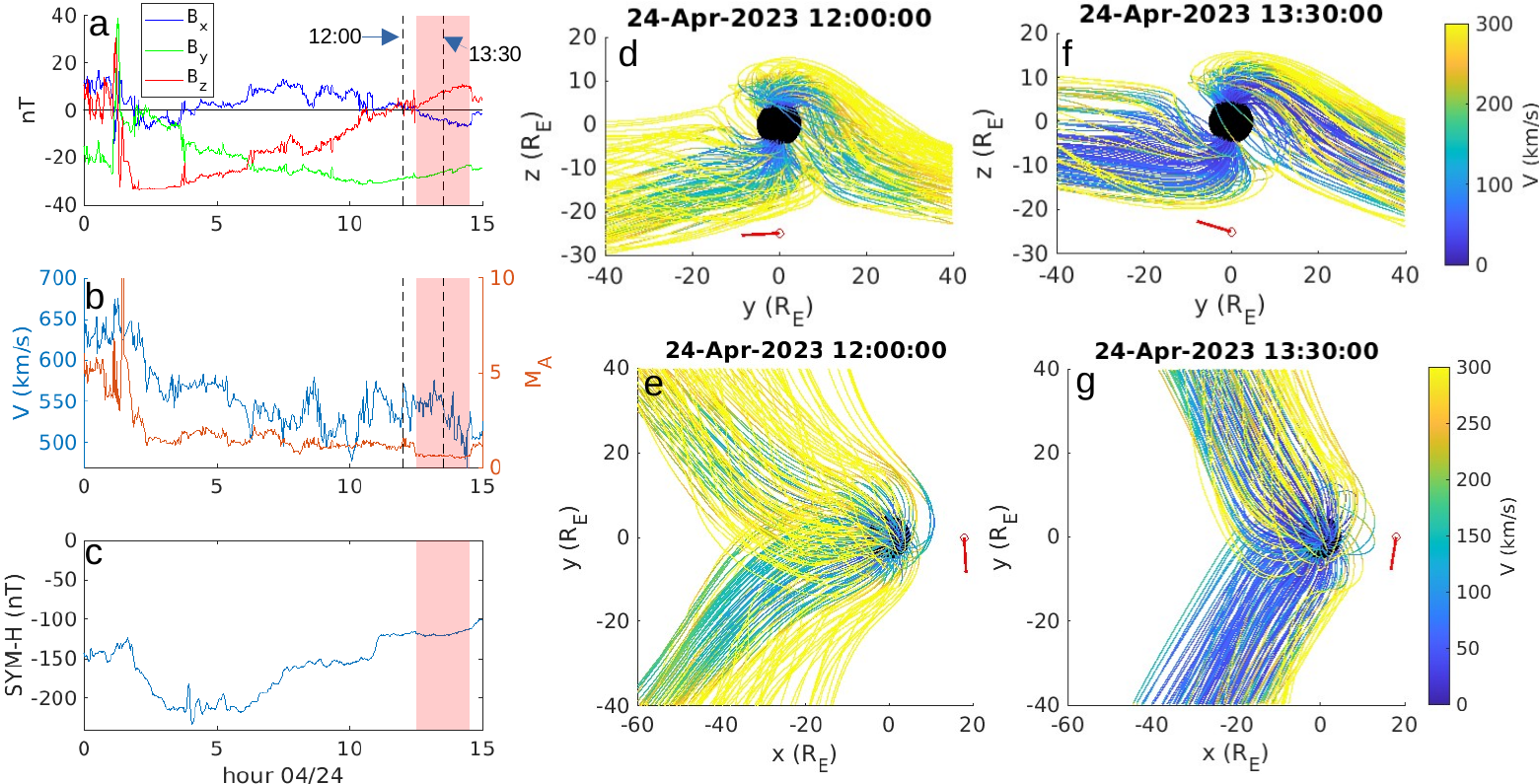}
	\caption{OMNI data on April 24, 2023: (a) IMF in SM coordinates, (b) plasma velocity (left axis) and M$_A$ (right axis), (c) storm-time SYM-H index. Red shading highlights the sub-Alfv\'{e}nic interval. Simulated 3D magnetic field line traces for polar cap fluxes before (d-e) and during (f-g) the Alfv\'{e}n wing interval. The y-z projections are shown in (d,f) and x-y projections in (e,g). Corresponding times are marked by vertical dashed lines in (a). Field line color represents local plasma speed and the red line shows upstream IMF orientation (strongly dawnward).}
	\label{fig:fig1}
\end{figure}

Simulations and observations of the April 2023 geomagnetic storm have revealed Earth's dynamic transformation to an AW configuration \cite{Burkholder2024,Chen2024,Gurram2025,Beedle2024}. Figures \ref{fig:fig1}a-b show the interplanetary magnetic field (IMF), solar wind speed (left side blue axis), and M$_A$ (right side orange axis) from NASA OMNIWeb \cite{omni1min} on April 24 (see longer interval in \citeA{Burkholder2024} Figure 1). The sub-Alfv\'{e}nic (M$_A\sim0.6$) interval begins at 12:30 UT and ends at 14:30 UT (red highlighting). During the AW interval, the IMF was primarily oriented along -B$_y$ ($\sim$-30 nT) with B$_z$ changing sign at the beginning then growing to $\sim$10 nT before the super-Alfv\'{e}nic solar wind returned. Figure \ref{fig:fig1}c shows SYM-H index with peak main phase value $\sim$-230 nT. The sub-Alfv\'{e}nic interval occurs during the recovery phase, corresponding to a SYM-H plateau at $\sim$-120 nT \cite{Chen2024}. 

Figures \ref{fig:fig1}d-g show simulated polar cap magnetic field lines at two selected times (vertical dashed lines in Figures \ref{fig:fig1}a-b): 12:00 UT (before the AW interval M$_A\sim1.4$), and 13:30 UT (in the middle of the AW interval M$_A\sim0.6$). The y-z views (Figures \ref{fig:fig1}d and f) show northern polar cap field lines map to the dusk and southern polar cap field lines map to the dawn (due to -B$_y$ dominant IMF, see red line illustrating the IMF orientation). Field line colors represent local plasma flow speed, showing many AW field lines (Figure \ref{fig:fig1}f) have slower moving plasma than their super-Alfv\'{e}nic counterpart (lobe magnetic field lines, Figure \ref{fig:fig1}d). Note, solar wind flow speed (Figure \ref{fig:fig1}b) is roughly equal at 12:00 and 13:30, indicating the simulated flow disturbances result from AW physics \cite{Ridley2007}. Faster flows ($>$200 km/s) are limited to the AW edges, a typical AW characteristic that has been identified in simulations and observations \cite{Chane2012,Chane2015}. Figures \ref{fig:fig1}e and g show the same magnetic field lines viewed from above the x-y plane. AW field lines (Figure \ref{fig:fig1}g) map immediately in the IMF direction after coming out of the polar cap (more accurately, they are tilted with respect to the upstream IMF by $\tan^{-1}$(M$_A$)), while lobe fluxes (Figure \ref{fig:fig1}e) in the near-planet magnetotail have a significant B$_x$ component. Only after mapping 10s of R$_E$ down-tail do those lobe field lines bend towards the IMF direction.

\section{Alfv\'{e}n Wing Field-Aligned Currents and Aurora}

\begin{sidewaysfigure}
	\centering
	\includegraphics[width=\linewidth]{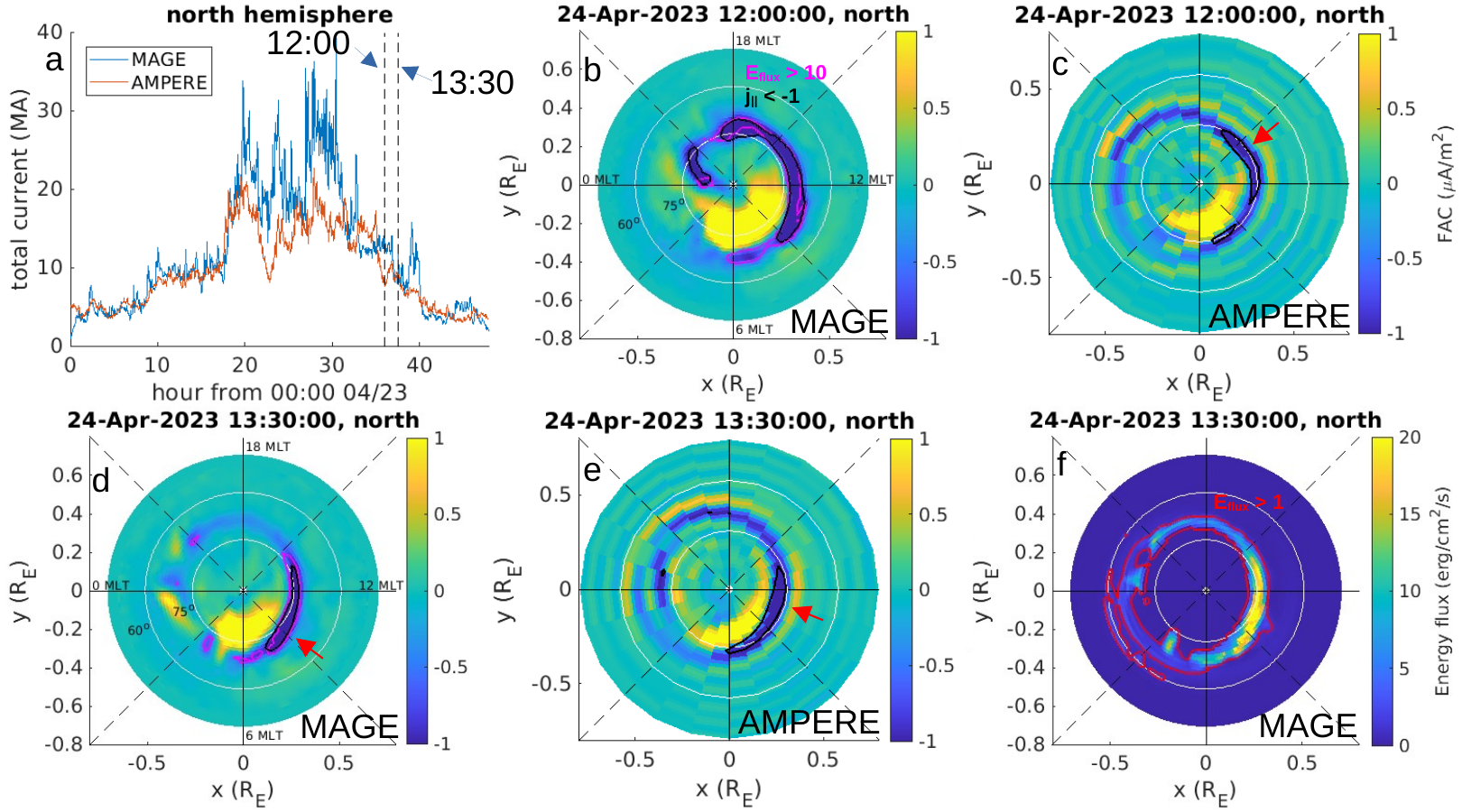}
	\caption{(a) Simulated (blue) and observed (orange) northern hemisphere total current. Vertical dashed lines correspond to the times from Figure \ref{fig:fig1}. (b) MAGE and (c) AMPERE northern hemisphere FAC, pre-Alfv\'{e}n wing. (d) MAGE and (e) AMPERE northern hemisphere FAC during the Alfv\'{e}n wing interval. Black contours enclose $j_{||}<-1$ $\mu$A/m$^2$. (f) MAGE precipitating electron energy flux (E$_{flux}$, a proxy for the aurora) during the Alfv\'{e}n wing interval. Red contour outlines the auroral oval (E$_{flux}>1$ erg/cm$^2$/s). The pink contour in (b) and (d) encloses E$_{flux}>10$ erg/cm$^2$/s.}
	\label{fig:fig2}
\end{sidewaysfigure}

Figure \ref{fig:fig2}a shows northern hemisphere total current ($I_{tot} = \frac{r^2}{2}\int\int|j_{||}|\sin\theta d\theta d\phi$ with $d\theta$ and $d\phi$ representing integration over MLat and MLT, respectively) from MAGE (blue) and AMPERE (orange) for the entire storm. The CME impact occurred around 17:45 UT on April 23. The main features of AMPERE $I_{tot}$ are reproduced by MAGE, including the quiet period before the storm, sharp increase of total current at the CME impact, and maxima around 21:00 UT April 23 and 3:00 UT April 24 (27 on the hours axis). Simulated $I_{tot}$ is overall slightly larger than observed $I_{tot}$ during the storm main phase, but decreases at a similar rate during the storm recovery, which began at $\sim$6:00 UT April 24 (30 on the hours axis). Vertical dashed lines mark 12:00 and 13:30 UT April 24 (36 and 37.5 on the hours axis), as in Figure \ref{fig:fig1}. Simulated and observed $I_{tot}$ from 12:00-15:00 UT April 24 (36-39 on the hours axis) are both $\sim10$ MA and relatively steady compared to the rest of the storm. This agreement provides a first point of validation that simulated FAC profiles are physically realistic during the AW interval.

Figures \ref{fig:fig2}b and c show MAGE and AMPERE northern hemisphere FAC, respectively, at 12:00 UT April 24 (a movie is included in the supplementary material). MAGE FAC is plotted at 120 km altitude, while AMPERE is at 780 km (both AMPERE and MAGE are plotted from 45-90$^\circ$ MLat). Strong downward (positive) FAC (yellow) is localized from 3-12 MLT for both simulation and observation. In the simulation, upward (negative) FAC  reaches $j_{||}<-1$ $\mu$A/m$^2$ (black contour) for MLTs $\sim$9-18, while AMPERE $j_{||}<-1$ $\mu$A/m$^2$ for MLT $\sim$11-16 (see red arrow). Figures \ref{fig:fig2}d and e show simulated and observed northern hemisphere FAC, respectively, at 13:30 UT April 24. Downward FAC is similar to the pre-AW interval, but not extending below 75$^\circ$ MLat. The $j_{||}<-1$ $\mu$A/m$^2$ black contour is shifted pre-noon during the AW interval (centered at MLT $\sim$8-10 for MAGE and MLT $\sim$10-11 for AMPERE, see red arrows). In the simulation, upward FACs at midnight and in the MLT 15-18 sector are suppressed at 13:30 UT compared to 12:00 UT. The MLT 15-18 sector is also suppressed according to AMPERE, but strong reduction of midnight FAC is not observed, although, upward FACs at midnight are shifted poleward. Figure \ref{fig:fig2}f shows MAGE precipitating electron energy flux (E$_{flux}$) at 13:30 UT. The auroral oval in this study is identified as E$_{flux} >$1 erg/cm$^2$/s \cite{Sorathia2024}, indicated by the red contour. The pink contour in Figures \ref{fig:fig2}b and d shows E$_{flux} >10$ erg/cm$^2$/s, which is mostly overlapping the black contour, indicating that intense E$_{flux}$ within the auroral oval corresponds to intense upward FAC. Furthermore, greater E$_{flux}$ roughly corresponds to brighter aurora, neglecting atmospheric chemistry. Thus, using E$_{flux}$ as the proxy for auroral emission, the northern hemisphere AW-associated aurora was likely most intense in the dawn day-side sector (MLT 8-13) during this event.

\begin{figure}
	\centering
	\includegraphics[width=\linewidth]{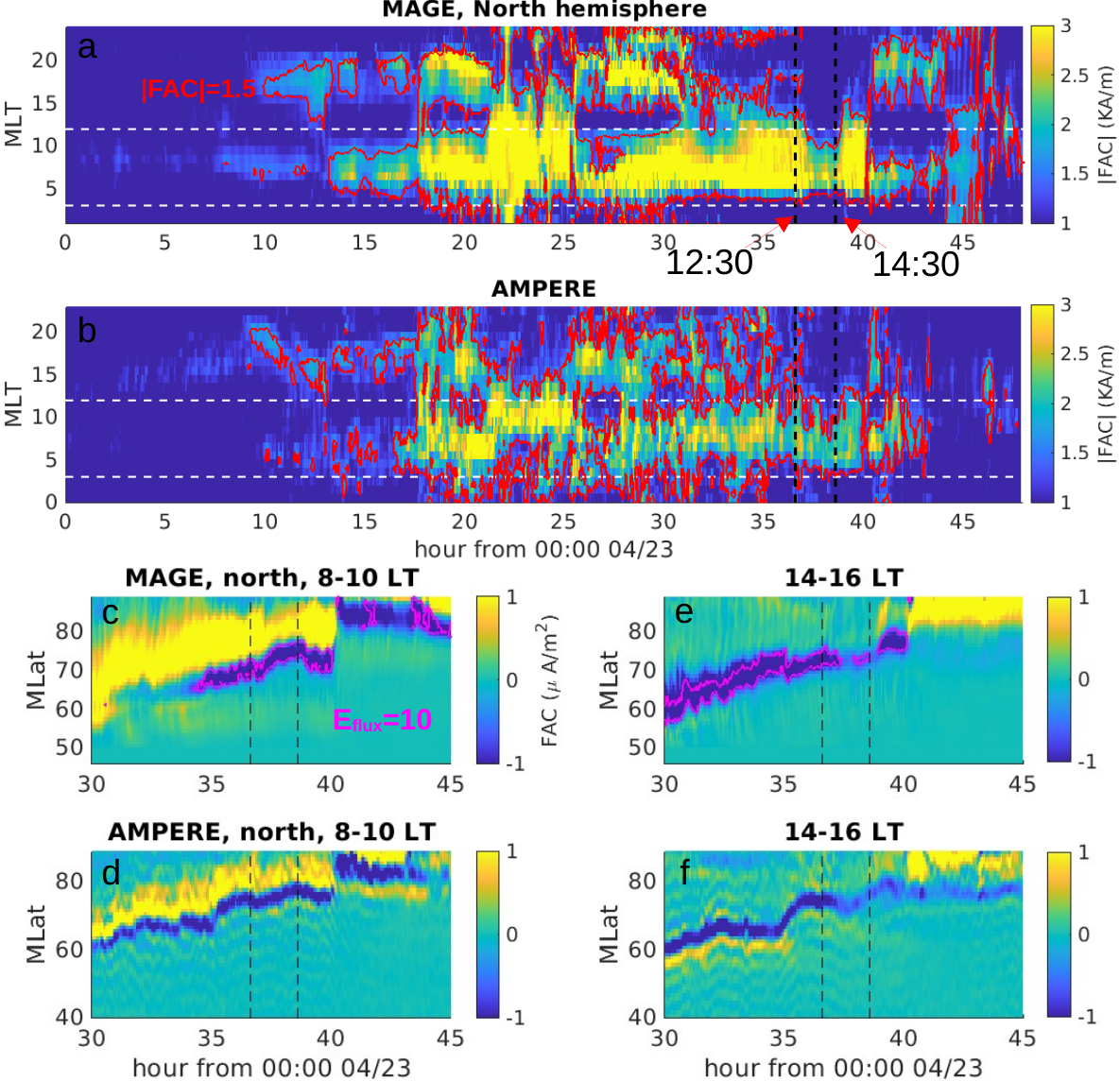}
	\caption{(a) MAGE and (b) AMPERE total current as a function of MLT ($r\int|j_{||}|d\theta$ in 1-hour MLT bins) for the duration of the storm. Red contour outlines $r\int|j_{||}|d\theta > 1.5$ KA/m. (c) MAGE and (d) AMPERE FAC MLat keogram averaged over the MLT sector 8-10. (e) MAGE and (f) AMPERE FAC MLat keogram averaged over the MLT sector 14-16.MLat keograms are zoomed-in to the period surrounding the sub-Alfv\'{e}nic interval (between dashed lines). Pink contours in (c) and (e) outline E$_{flux}>10$ erg/cm$^2$/s.}
	\label{fig:fig3}
\end{figure}

Figures \ref{fig:fig3}a and b compare simulated and observed FAC as a function of MLT for the entire storm. The quantity $r\int|j_{||}|d\theta$ is summed over 1-hour MLT bins for MAGE (Figure \ref{fig:fig3}a) and AMPERE (Figure \ref{fig:fig3}b). Vertical dashed lines represent the start (12:30) and end (14:30) of sub-Alfv\'{e}nic solar wind. Over the course of the storm, the simulated FAC distribution across MLT agrees well with observations. In particular, the AW transformation is characterized by $r\int|j_{||}|d\theta > 1.5$ KA/m (red contour) being isolated to the MLT sector 3-12 (white dashed lines). Figures \ref{fig:fig3}c (MAGE) and d (AMPERE) show MLat keogram representations (constant slices of MLT stacked left to right) of FAC averaged in the MLT sector 8-10 (day-side dawn sector, the location where Figures \ref{fig:fig2}d and e show strong upward FAC). During the AW interval (between vertical dashed lines), upward FAC in the 8-10 MLT sector is localized around 70$^\circ$ MLat. Figures \ref{fig:fig3}e (MAGE) and f (AMPERE) show MLat keogram representations of FAC averaged over the MLT sector 14-16 (day-side dusk sector, the location where Figures \ref{fig:fig2}d and e show weak upward FAC). During the AW interval, upward FACs in the 14-16 MLT sector are noticeably suppressed.The pink contour in Figures \ref{fig:fig3}c and e outlines  E$_{flux}>10$ erg/cm$^2$/s, showing the simulated auroral proxy has the same distribution as upward FAC. Notice, the time delay between the beginning of the sub-Alfv\'{e}nic interval and the suppression of upward FAC in the 14-16 MLT sector is similar for MAGE and AMPERE.

\begin{sidewaysfigure}
	\centering
	\includegraphics[width=\linewidth]{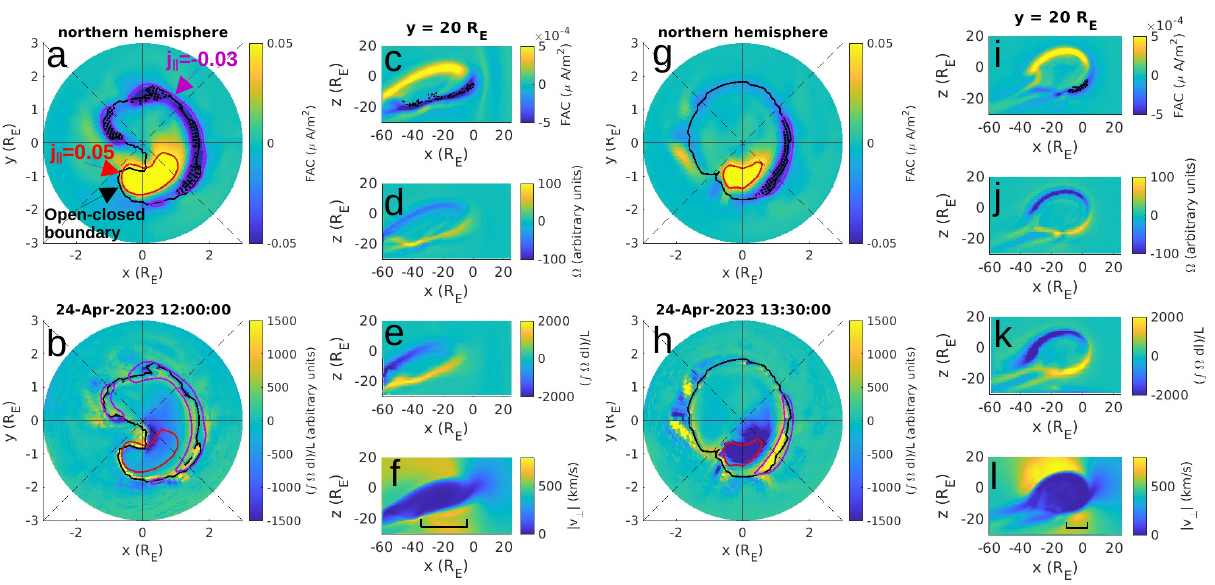}
	\caption{(a) Simulated northern hemisphere FAC, pre-Alfv\'{e}n wing (12:00 UT), with black (open-closed boundary), red ($j_{||} > 0.05 \mu$A/m$^2$), and purple ($j_{||}< -0.03 \mu$A/m$^2$) contours. Black dots are the start location of field line traces (at r = 3 R$_E$). (b) Simulated magnetic field line integrated $\mathcal{I} = L^{-1}\int\Omega d\ell$ with same contours. Simulation slices at y = 20 R$_E$: (c) FAC with black dots showing the field line locations from (a) at y = 20 R$_E$, (d) $\Omega$, (e) $\mathcal{I}$ (integrated from y = 20 to y = 60 R$_E$), (f) magnetic field perpendicular velocity $v_\perp$ (black bar highlights the flow shear region at the southern edge of the north lobe flux). (g-l) The same panels during the Alfv\'{e}n wing interval (13:30 UT).}
	\label{fig:fig4}
\end{sidewaysfigure}

Figure \ref{fig:fig4}a shows northern hemisphere FAC pre-AW (12:00 UT), the same as Figure \ref{fig:fig2}b, except on a sphere with $r = 3$ R$_E$ (for field line tracing within the MHD grid). Strong downward and upward FAC are outlined with red ($j_{||} >$ 0.05 $\mu A/m^2$) and purple ($j_{||} <$ -0.03 $\mu A/m^2$), respectively. The black contour is the open-closed magnetic field boundary, showing the downward FAC enhancement lies mostly on open polar cap fluxes, while the upward FAC enhancement straddles the day-side open-closed boundary.

Equations 1 and 2 from \citeA{Liu2022} describe FAC contributions from the diamagnetic current and change in magnetic field-aligned flow vorticity $\Omega = (\nabla \times \mathbf{V})\cdot\hat{\mathbf{b}}$, respectively. Simulated pressure gradients contribute an order of magnitude less FAC than velocity shear, so the diamagnetic component, which depends on thermal and magnetic pressure gradients, is neglected. The inertial component is (\citeA{Liu2022} Equation 2): 

$$j_{||} = \int \frac{\rho}{B}\frac{d\Omega}{dt}d\ell,$$

\noindent where $d\ell$ represents integration along a magnetic field line. The inertial component is a FAC driven by $\frac{d\Omega}{dt}$ \cite{Hasegawa1979}. To isolate the vorticity contribution to FAC generation, Figure \ref{fig:fig4}b shows $\mathcal{I} = L^{-1}\int\Omega d\ell$ integrated along magnetic field lines from r = 3 R$_E$ to the edge of a 60 R$_E$ cube (the factor $L^{-1}$ normalizes by the field line length), during the pre-AW interval. Within red and purple contours (which are the same as Figure \ref{fig:fig4}a), $\mathcal{I}$ is slightly enhanced above the background, but the greatest values of $\mathcal{I}$ are in the midnight MLT sector, associated with the ``bite-out'' giving the polar cap a kidney bean shape. This feature of the open-closed boundary is due to Dungey-type magnetotail reconnection that removes lobe flux from the polar caps and adds it to the magnetosphere.

Figures \ref{fig:fig4}c-f are 2D visualizations at y = 20 R$_E$, a plane which intersects north lobe field lines extending duskward (see Figure \ref{fig:fig1}d). Figure \ref{fig:fig4}c shows FAC, with black dots indicating magnetic connection to the same black dots in Figure \ref{fig:fig4}a (demonstrating where large upward FAC in the northern polar cap maps to the y = 20 R$_E$ plane). Those strong upward FACs in the northern polar cap map to the southern edge of the dusk lobe. Correspondingly, the northern hemisphere downward FAC maps to the northern edge of the dusk lobe. Figure \ref{fig:fig4}d shows $\Omega$, which is localized to the same locations as FAC. Given the relatively stable solar wind driving before (and during) the AW interval, the $\partial/\partial t$ portion of $d\Omega/dt$ contributes relatively little to FAC generation. The spatial gradient of $\Omega$, which is large because of strong localization at the lobe boundary, is driving the FACs. Figure \ref{fig:fig4}e shows $\mathcal{I}$ integrated from y = 20 R$_E$ to 60 R$_E$. The enhancements of $\mathcal{I}$ are similar to the local value of $\Omega$ (Figure \ref{fig:fig4}d). Peak positive $\mathcal{I}$ is at x $\sim-40$ R$_E$ and the positive enhancement extends from $x = -60$ R$_E$ to -10 R$_E$. Figure \ref{fig:fig4}f shows the magnetic field perpendicular velocity ($v_\perp$), demonstrating the velocity shear between tailward moving magnetosheath magnetic field and slower convecting lobe fluxes. The sharp boundary of $v_\perp$ localizes $\Omega$ because magnetic field lines are twisted only in a thin shear layer.

Figures \ref{fig:fig4}g-l are the same as Figures \ref{fig:fig4}a-f, but for the AWs (13:30 UT). Figure \ref{fig:fig4}g shows the open-closed boundary ``bite-out,'' associated with night-side AW reconnection, is at 3-4 MLT \cite{Burkholder2024}. Similar to Figure \ref{fig:fig4}a, the downward FAC enhancement in Figure \ref{fig:fig4}g lies adjacent to the bite-out and upward FAC straddles the day-side open-closed boundary. Figure \ref{fig:fig4}h demonstrates the AW FAC structures (red and purple contours) are spatially co-located with the strongest enhancements of $|\mathcal{I}|$, different than the pre-AW interval. Downward (positive) FAC corresponds to large negative $\mathcal{I}$ and upward (negative) FAC to large positive $\mathcal{I}$.

Figure \ref{fig:fig4}i shows magnetic field lines seeded at the black points in Figure \ref{fig:fig4}g map to a relatively localized (compared to Figure \ref{fig:fig4}c) portion of the southern edge of the dusk AW. The cross-sectional shape of the AW is also significantly different than the pre-AW interval. The ``wing'' is more circular compared to lobe fluxes during the earlier interval, which actually resemble an aerodynamic wing  (e.g., Figure \ref{fig:fig4}c). Figure \ref{fig:fig4}j shows negative FAC on the southern edge of the dusk AW is associated with strong positive $\Omega$. Compared to pre-AW, local values of $\Omega$ are larger, especially at the northern edge of the wing. When $\Omega$ is integrated along the AW ($\mathcal{I}$, Figure \ref{fig:fig4}k), there are layers of strong vorticity along the northern and southern edges of the wing. Peak positive $\mathcal{I}$ is at x $\sim-10$ R$_E$ and strong negative $\mathcal{I}$ (on the northern edge of the dusk wing) extends from x $\sim-45$ to -5 R$_E$. Figure \ref{fig:fig4}l shows velocity shear on the northern part of the wing (mapping within the red contour, Figure \ref{fig:fig4}g) is greater at 13:30 compared to 12:00, which is due to the lack of a bow shock during the sub-Alfv\'{e}nic solar wind, and because AW structures are more inclined to the oncoming V$_{sw}$ compared to typical magnetotail lobes (the wing-V$_{sw}$ angle is larger than the lobe-V$_{sw}$ angle). The southern edge of the AW has only a small extent 10-15 R$_E$ along the x-direction with significant velocity shear and corresponding negative FAC generation (compare black bar in Figures \ref{fig:fig4}f and l), such that simulated northern hemisphere AW E$_{flux}$ is localized in the dawn day-side sector (see Figure \ref{fig:fig2}f) during this event.

\section{Summary and Conclusions}
AWs are a unique magnetospheric configuration where the bow shock disappears and night-side lobe fluxes separate. Unshocked solar wind impinging directly on Earth's magnetic field leads to greater velocity shear between the IMF and relatively stationary AW fluxes. Localized vorticity drives FACs at the AW edges which flow towards the ionosphere as part of the Birkeland currents. The main feature of northern hemisphere simulated AW auroral precipitation with dawnward (-B$_y$) IMF orientation is localization in the MLT sector 8-13. Just before the AWs formed, the IMF was also dawnward and M$_A \sim 1.4$, but northern hemisphere upward FAC extended from 9-18 MLT. The different extent of these upward FAC regions corresponds to the different shape of the southern edge of the dusk AW compared to the corresponding lobe fluxes. During sub-Alfv\'{e}nic driving, the AW cross-section is roughly circular, whereas the equivalent shape is flattened for lobe fluxes during super-Alfv\'{e}nic solar wind (note, the circular AW cross-section is likely not unique to B$_y$ dominant IMF). Flow vorticity at the dusk AW southern edge is localized between x = -10 and 5 R$_E$, which maps to the relatively small region of MLT 8-13 in the northern polar cap, while velocity shear at the southern edge of the magnetospheric lobe fluxes extends from x = -45 to -5 R$_E$, and correspondingly maps to a larger MLT sector. Simulations of AW auroral morphology under different IMF orientations are left for future work.

MAGE reproduces northern hemisphere FAC observed by AMPERE remarkably well during the April 2023 storm (see Figures \ref{fig:fig2}a and \ref{fig:fig3}). Furthermore, simulated northern and southern (not shown) hemisphere FACs are essentially (mirror) symmetric (although weaker in the south). However, the southern hemisphere observed by AMPERE (see AMPERE quicklook plots at https://ampere.jhuapl.edu/products/) is significantly different, indicating a stronger response to night-side dynamics. As mentioned in Section 2, this discrepancy could be a data processing artifact, but the simulation also indicates cusp reconnection is less active in the southern hemisphere. The physical mechanism driving this interhemispheric asymmetry is left for future work. Also, while MAGE results agree well with AMPERE during the AW interval, incorporating Alfv\'{e}nic aurora in MAGE may also improve the consistency between the model and data. Alfv\'{e}n waves can be a major contributor to electron acceleration during geomagnetic storms \cite{Keiling2019}, though the expected contribution during the transition to AWs requires further exploration.

Unraveling unique behavior of the Earth's AWs can help better understand analogous systems, like plasma experiments operating in the sub-Alfv\'{e}nic flow regime, and outer planet moons which are embedded either intermittently or continuously in sub-Alfv\'{e}nic plasmas. Europa, soon to be visited by JUICE and Europa Clipper, is a high priority science target among such moons. Despite not having an intrinsic magnetic field, Europa has AWs which are connected to Jupiter. On those magnetic field lines, energetic particles can more easily access the surface, an important consideration in the search for signs of life \cite{Meitzler2023}. Simulations and observations of Earth's AWs can help to anticipate possible signatures of Europa's AWs before JUICE and Europa Clipper arrive in the 2030s. In addition, some exoplanets orbit within the Alfv\'{e}n zone of their host star, leaving them in a perpetual state of AWs. The aurora on such worlds may be similar to the Earth's AWs.

\section{Open Research}
The MAGE simulation outputs are archived via \cite{Burkholder2025b}. AMPERE data are available from JHU/APL at \url{https://ampere.jhuapl.edu/}.
%


\acknowledgments
We acknowledge high-performance computing support from the Derecho system (doi:10.5065/qx9a-pg09) provided by the NSF National Center for Atmospheric Research (NCAR), sponsored by the National Science Foundation. We acknowledge use of NASA/GSFC’s Space Physics Data Facility’s OMNIWeb (https://omniweb.gsfc.nasa.gov/). We thank the team of the Center for Geospace Storms for providing the MAGE model. We thank the AMPERE team and the AMPERE Science Data Center for providing data products derived from the Iridium Communications constellation, enabled by support from the National Science Foundation. Funding for this work is provided by the Partnership for Heliospheric and Space Environment Research (PHaSER) and MMS Early Career Grant 80NSSC25K7352. D.L. and K.S. were supported by NASA DRIVE Science Center for Geospace Storms (CGS) under Grant 80NSSC22M0163. D. L. was also supported by Grants 80NSSC23K1055, 80NSSC22K1635, 80NSSC21K1677, and 80NSSC21K0008. Work at DIAS was supported by Taighde \'{E}ireann - Research Ireland Laureate Consolidator award SOLMEX (CFB).



%
\bibliography{AMPERE_REMIX_alfwings}

\begin{thebibliography}{}

\bibitem [\protect \citeauthoryear {%
Anderson%
\ \protect \BOthers {.}}{%
Anderson%
\ \protect \BOthers {.}}{%
{\protect \APACyear {2021}}%
}]{%
Anderson2021}
\APACinsertmetastar {%
Anderson2021}%
\begin{APACrefauthors}%
Anderson, B\BPBI J.%
, Angappan, R.%
, Barik, A.%
, Vines, S\BPBI K.%
, Stanley, S.%
, Bernasconi, P\BPBI N.%
\BDBL {}Barnes, R\BPBI J.%
\end{APACrefauthors}%
\unskip\
\newblock
\APACrefYearMonthDay{2021}{}{}.
\newblock
{\BBOQ}\APACrefatitle {Iridium Communications Satellite Constellation Data for
  Study of Earth's Magnetic Field} {Iridium communications satellite
  constellation data for study of earth's magnetic field}.{\BBCQ}
\newblock
\APACjournalVolNumPages{Geochemistry, Geophysics,
  Geosystems}{22}{8}{e2020GC009515}.
\newblock
\begin{APACrefDOI} \doi{https://doi.org/10.1029/2020GC009515} \end{APACrefDOI}
\PrintBackRefs{\CurrentBib}

\bibitem [\protect \citeauthoryear {%
Anderson%
\ \protect \BOthers {.}}{%
Anderson%
\ \protect \BOthers {.}}{%
{\protect \APACyear {2014}}%
}]{%
Anderson2014}
\APACinsertmetastar {%
Anderson2014}%
\begin{APACrefauthors}%
Anderson, B\BPBI J.%
, Korth, H.%
, Waters, C\BPBI L.%
, Green, D\BPBI L.%
, Merkin, V\BPBI G.%
, Barnes, R\BPBI J.%
\BCBL {}\ \BBA {} Dyrud, L\BPBI P.%
\end{APACrefauthors}%
\unskip\
\newblock
\APACrefYearMonthDay{2014}{}{}.
\newblock
{\BBOQ}\APACrefatitle {Development of large-scale Birkeland currents determined
  from the Active Magnetosphere and Planetary Electrodynamics Response
  Experiment} {Development of large-scale birkeland currents determined from
  the active magnetosphere and planetary electrodynamics response
  experiment}.{\BBCQ}
\newblock
\APACjournalVolNumPages{Geophysical Research Letters}{41}{9}{3017-3025}.
\newblock
\begin{APACrefDOI} \doi{https://doi.org/10.1002/2014GL059941} \end{APACrefDOI}
\PrintBackRefs{\CurrentBib}

\bibitem [\protect \citeauthoryear {%
Anderson%
, Takahashi%
, Kamei%
, Waters%
\BCBL {}\ \BBA {} Toth%
}{%
Anderson%
\ \protect \BOthers {.}}{%
{\protect \APACyear {2002}}%
}]{%
Anderson2002}
\APACinsertmetastar {%
Anderson2002}%
\begin{APACrefauthors}%
Anderson, B\BPBI J.%
, Takahashi, K.%
, Kamei, T.%
, Waters, C\BPBI L.%
\BCBL {}\ \BBA {} Toth, B\BPBI A.%
\end{APACrefauthors}%
\unskip\
\newblock
\APACrefYearMonthDay{2002}{}{}.
\newblock
{\BBOQ}\APACrefatitle {Birkeland current system key parameters derived from
  Iridium observations: Method and initial validation results} {Birkeland
  current system key parameters derived from iridium observations: Method and
  initial validation results}.{\BBCQ}
\newblock
\APACjournalVolNumPages{Journal of Geophysical Research: Space
  Physics}{107}{A6}{SMP 11-1-SMP 11-13}.
\newblock
\begin{APACrefDOI} \doi{https://doi.org/10.1029/2001JA000080} \end{APACrefDOI}
\PrintBackRefs{\CurrentBib}

\bibitem [\protect \citeauthoryear {%
Andersson%
\ \BBA {} Ergun%
}{%
Andersson%
\ \BBA {} Ergun%
}{%
{\protect \APACyear {2012}}%
}]{%
Andersson2012}
\APACinsertmetastar {%
Andersson2012}%
\begin{APACrefauthors}%
Andersson, L.%
\BCBT {}\ \BBA {} Ergun, R.%
\end{APACrefauthors}%
\unskip\
\newblock
\APACrefYearMonthDay{2012}{}{}.
\newblock
{\BBOQ}\APACrefatitle {The search for double layers in space plasmas} {The
  search for double layers in space plasmas}.{\BBCQ}
\newblock
\APACjournalVolNumPages{Auroral phenomenology and magnetospheric processes:
  Earth And other planets}{197}{}{241--250}.
\PrintBackRefs{\CurrentBib}

\bibitem [\protect \citeauthoryear {%
Beedle%
\ \protect \BOthers {.}}{%
Beedle%
\ \protect \BOthers {.}}{%
{\protect \APACyear {2024}}%
}]{%
Beedle2024}
\APACinsertmetastar {%
Beedle2024}%
\begin{APACrefauthors}%
Beedle, J\BPBI M\BPBI H.%
, Chen, L\BHBI J.%
, Shuster, J\BPBI R.%
, Gurram, H.%
, Gershman, D\BPBI J.%
, Chen, Y.%
\BDBL {}Torbert, R\BPBI B.%
\end{APACrefauthors}%
\unskip\
\newblock
\APACrefYearMonthDay{2024}{}{}.
\newblock
{\BBOQ}\APACrefatitle {Field-Aligned Current Structures During the Terrestrial
  Magnetosphere's Transformation Into Alfv\'{e}n Wings and Recovery}
  {Field-aligned current structures during the terrestrial magnetosphere's
  transformation into alfv\'{e}n wings and recovery}.{\BBCQ}
\newblock
\APACjournalVolNumPages{Geophysical Research Letters}{51}{13}{e2024GL108839}.
\newblock
\begin{APACrefDOI} \doi{https://doi.org/10.1029/2024GL108839} \end{APACrefDOI}
\PrintBackRefs{\CurrentBib}

\bibitem [\protect \citeauthoryear {%
Birkeland%
}{%
Birkeland%
}{%
{\protect \APACyear {1908}}%
}]{%
Birkeland1908}
\APACinsertmetastar {%
Birkeland1908}%
\begin{APACrefauthors}%
Birkeland, K.%
\end{APACrefauthors}%
\unskip\
\newblock
\APACrefYear{1908}.
\newblock
\APACrefbtitle {The Norwegian aurora polaris expedition, 1902-1903} {The
  norwegian aurora polaris expedition, 1902-1903}\ (\BVOL~1).
\newblock
\APACaddressPublisher{}{Christiania, H. Aschelhoug}.
\PrintBackRefs{\CurrentBib}

\bibitem [\protect \citeauthoryear {%
Birkeland%
}{%
Birkeland%
}{%
{\protect \APACyear {1913}}%
}]{%
Birkeland1913}
\APACinsertmetastar {%
Birkeland1913}%
\begin{APACrefauthors}%
Birkeland, K.%
\end{APACrefauthors}%
\unskip\
\newblock
\APACrefYear{1913}.
\newblock
\APACrefbtitle {The Norwegian aurora polaris expedition, 1902-1903} {The
  norwegian aurora polaris expedition, 1902-1903}\ (\BVOL~2).
\newblock
\APACaddressPublisher{}{Christiania, H. Aschelhoug}.
\PrintBackRefs{\CurrentBib}

\bibitem [\protect \citeauthoryear {%
B.~Burkholder%
}{%
B.~Burkholder%
}{%
{\protect \APACyear {2025}}%
}]{%
Burkholder2025b}
\APACinsertmetastar {%
Burkholder2025b}%
\begin{APACrefauthors}%
Burkholder, B.%
\end{APACrefauthors}%
\unskip\
\newblock
\APACrefYearMonthDay{2025}{}{}.
\newblock
{\BBOQ}\APACrefatitle {MAGE Global Geospace: April 2023 storm [Data set]} {Mage
  global geospace: April 2023 storm [data set]}.{\BBCQ}
\newblock
\APACjournalVolNumPages{In Geophysical Research Letters. Zenodo}{}{}{}.
\newblock
\begin{APACrefDOI} \doi{https://doi.org/10.5281/zenodo.14990397}
  \end{APACrefDOI}
\PrintBackRefs{\CurrentBib}

\bibitem [\protect \citeauthoryear {%
B\BPBI L.~Burkholder%
\ \protect \BOthers {.}}{%
B\BPBI L.~Burkholder%
\ \protect \BOthers {.}}{%
{\protect \APACyear {2024}}%
}]{%
Burkholder2024}
\APACinsertmetastar {%
Burkholder2024}%
\begin{APACrefauthors}%
Burkholder, B\BPBI L.%
, Chen, L\BHBI J.%
, Sarantos, M.%
, Gershman, D\BPBI J.%
, Argall, M\BPBI R.%
, Chen, Y.%
\BDBL {}Gurram, H.%
\end{APACrefauthors}%
\unskip\
\newblock
\APACrefYearMonthDay{2024}{}{}.
\newblock
{\BBOQ}\APACrefatitle {Global Magnetic Reconnection During Sustained
  Sub-Alfv\'{e}nic Solar Wind Driving} {Global magnetic reconnection during
  sustained sub-alfv\'{e}nic solar wind driving}.{\BBCQ}
\newblock
\APACjournalVolNumPages{Geophysical Research Letters}{51}{6}{e2024GL108311}.
\newblock
\begin{APACrefDOI} \doi{https://doi.org/10.1029/2024GL108311} \end{APACrefDOI}
\PrintBackRefs{\CurrentBib}

\bibitem [\protect \citeauthoryear {%
Chan\'{e}%
\ \protect \BOthers {.}}{%
Chan\'{e}%
\ \protect \BOthers {.}}{%
{\protect \APACyear {2015}}%
}]{%
Chane2015}
\APACinsertmetastar {%
Chane2015}%
\begin{APACrefauthors}%
Chan\'{e}, E.%
, Raeder, J.%
, Saur, J.%
, Neubauer, F\BPBI M.%
, Maynard, K\BPBI M.%
\BCBL {}\ \BBA {} Poedts, S.%
\end{APACrefauthors}%
\unskip\
\newblock
\APACrefYearMonthDay{2015}{}{}.
\newblock
{\BBOQ}\APACrefatitle {Simulations of the Earth's magnetosphere embedded in
  sub-Alfv\'{e}nic solar wind on 24 and 25 May 2002} {Simulations of the
  earth's magnetosphere embedded in sub-alfv\'{e}nic solar wind on 24 and 25
  may 2002}.{\BBCQ}
\newblock
\APACjournalVolNumPages{Journal of Geophysical Research: Space
  Physics}{120}{10}{8517-8528}.
\newblock
\begin{APACrefDOI} \doi{https://doi.org/10.1002/2015JA021515} \end{APACrefDOI}
\PrintBackRefs{\CurrentBib}

\bibitem [\protect \citeauthoryear {%
Chan\'{e}%
, Saur%
, Neubauer%
, Raeder%
\BCBL {}\ \BBA {} Poedts%
}{%
Chan\'{e}%
\ \protect \BOthers {.}}{%
{\protect \APACyear {2012}}%
}]{%
Chane2012}
\APACinsertmetastar {%
Chane2012}%
\begin{APACrefauthors}%
Chan\'{e}, E.%
, Saur, J.%
, Neubauer, F\BPBI M.%
, Raeder, J.%
\BCBL {}\ \BBA {} Poedts, S.%
\end{APACrefauthors}%
\unskip\
\newblock
\APACrefYearMonthDay{2012}{}{}.
\newblock
{\BBOQ}\APACrefatitle {Observational evidence of Alfv\'{e}n wings at the Earth}
  {Observational evidence of alfv\'{e}n wings at the earth}.{\BBCQ}
\newblock
\APACjournalVolNumPages{Journal of Geophysical Research: Space
  Physics}{117}{A9}{}.
\newblock
\begin{APACrefDOI} \doi{https://doi.org/10.1029/2012JA017628} \end{APACrefDOI}
\PrintBackRefs{\CurrentBib}

\bibitem [\protect \citeauthoryear {%
Chaston%
\ \protect \BOthers {.}}{%
Chaston%
\ \protect \BOthers {.}}{%
{\protect \APACyear {2004}}%
}]{%
Chaston2004}
\APACinsertmetastar {%
Chaston2004}%
\begin{APACrefauthors}%
Chaston, C\BPBI C.%
, Bonnell, J\BPBI W.%
, Carlson, C\BPBI W.%
, McFadden, J\BPBI P.%
, Ergun, R\BPBI E.%
, Strangeway, R\BPBI J.%
\BCBL {}\ \BBA {} Lund, E\BPBI J.%
\end{APACrefauthors}%
\unskip\
\newblock
\APACrefYearMonthDay{2004}{}{}.
\newblock
{\BBOQ}\APACrefatitle {Auroral ion acceleration in dispersive Alfv\'{e}n waves}
  {Auroral ion acceleration in dispersive alfv\'{e}n waves}.{\BBCQ}
\newblock
\APACjournalVolNumPages{Journal of Geophysical Research: Space
  Physics}{109}{A4}{}.
\newblock
\begin{APACrefDOI} \doi{https://doi.org/10.1029/2003JA010053} \end{APACrefDOI}
\PrintBackRefs{\CurrentBib}

\bibitem [\protect \citeauthoryear {%
Chen%
\ \protect \BOthers {.}}{%
Chen%
\ \protect \BOthers {.}}{%
{\protect \APACyear {2024}}%
}]{%
Chen2024}
\APACinsertmetastar {%
Chen2024}%
\begin{APACrefauthors}%
Chen, L\BHBI J.%
, Gershman, D.%
, Burkholder, B.%
, Chen, Y.%
, Sarantos, M.%
, Jian, L.%
\BDBL {}Burch, J.%
\end{APACrefauthors}%
\unskip\
\newblock
\APACrefYearMonthDay{2024}{}{}.
\newblock
{\BBOQ}\APACrefatitle {Earth's Alfv\'{e}n Wings Driven by the April 2023
  Coronal Mass Ejection} {Earth's alfv\'{e}n wings driven by the april 2023
  coronal mass ejection}.{\BBCQ}
\newblock
\APACjournalVolNumPages{Geophysical Research Letters}{51}{14}{e2024GL108894}.
\newblock
\begin{APACrefDOI} \doi{https://doi.org/10.1029/2024GL108894} \end{APACrefDOI}
\PrintBackRefs{\CurrentBib}

\bibitem [\protect \citeauthoryear {%
Coxon%
\ \protect \BOthers {.}}{%
Coxon%
\ \protect \BOthers {.}}{%
{\protect \APACyear {2023}}%
}]{%
Coxon2023}
\APACinsertmetastar {%
Coxon2023}%
\begin{APACrefauthors}%
Coxon, J\BPBI C.%
, Chisham, G.%
, Freeman, M\BPBI P.%
, Forsyth, C.%
, Walach, M\BHBI T.%
, Murphy, K\BPBI R.%
\BDBL {}Fogg, A\BPBI R.%
\end{APACrefauthors}%
\unskip\
\newblock
\APACrefYearMonthDay{2023}{}{}.
\newblock
{\BBOQ}\APACrefatitle {Extreme Birkeland Currents Are More Likely During
  Geomagnetic Storms on the Dayside of the Earth} {Extreme birkeland currents
  are more likely during geomagnetic storms on the dayside of the
  earth}.{\BBCQ}
\newblock
\APACjournalVolNumPages{Journal of Geophysical Research: Space
  Physics}{128}{12}{e2023JA031946}.
\newblock
\begin{APACrefDOI} \doi{https://doi.org/10.1029/2023JA031946} \end{APACrefDOI}
\PrintBackRefs{\CurrentBib}

\bibitem [\protect \citeauthoryear {%
Crowley%
}{%
Crowley%
}{%
{\protect \APACyear {1991}}%
}]{%
Crowley1991}
\APACinsertmetastar {%
Crowley1991}%
\begin{APACrefauthors}%
Crowley, G.%
\end{APACrefauthors}%
\unskip\
\newblock
\APACrefYearMonthDay{1991}{}{}.
\newblock
{\BBOQ}\APACrefatitle {Dynamics of the Earth's Thermosphere: A Review}
  {Dynamics of the earth's thermosphere: A review}.{\BBCQ}
\newblock
\APACjournalVolNumPages{Reviews of Geophysics}{29}{S2}{1143-1165}.
\newblock
\begin{APACrefURL}
  \url{https://agupubs.onlinelibrary.wiley.com/doi/abs/10.1002/rog.1991.29.s2.1143}
  \end{APACrefURL}
\newblock
\begin{APACrefDOI} \doi{https://doi.org/10.1002/rog.1991.29.s2.1143}
  \end{APACrefDOI}
\PrintBackRefs{\CurrentBib}

\bibitem [\protect \citeauthoryear {%
Gurram%
\ \protect \BOthers {.}}{%
Gurram%
\ \protect \BOthers {.}}{%
{\protect \APACyear {2025}}%
}]{%
Gurram2025}
\APACinsertmetastar {%
Gurram2025}%
\begin{APACrefauthors}%
Gurram, H.%
, Shuster, J\BPBI R.%
, Chen, L\BHBI J.%
, Hasegawa, H.%
, Denton, R\BPBI E.%
, Burkholder, B\BPBI L.%
\BDBL {}Burch, J.%
\end{APACrefauthors}%
\unskip\
\newblock
\APACrefYearMonthDay{2025}{}{}.
\newblock
{\BBOQ}\APACrefatitle {Earth's Alfv\'{e}n Wings: Unveiling Dynamic Variations
  of Field-Line Topologies With Electron Distributions} {Earth's alfv\'{e}n
  wings: Unveiling dynamic variations of field-line topologies with electron
  distributions}.{\BBCQ}
\newblock
\APACjournalVolNumPages{Geophysical Research Letters}{52}{3}{e2024GL111931}.
\newblock
\begin{APACrefDOI} \doi{https://doi.org/10.1029/2024GL111931} \end{APACrefDOI}
\PrintBackRefs{\CurrentBib}

\bibitem [\protect \citeauthoryear {%
Hajra%
\ \BBA {} Tsurutani%
}{%
Hajra%
\ \BBA {} Tsurutani%
}{%
{\protect \APACyear {2022}}%
}]{%
Hajra2022}
\APACinsertmetastar {%
Hajra2022}%
\begin{APACrefauthors}%
Hajra, R.%
\BCBT {}\ \BBA {} Tsurutani, B\BPBI T.%
\end{APACrefauthors}%
\unskip\
\newblock
\APACrefYearMonthDay{2022}{feb}{}.
\newblock
{\BBOQ}\APACrefatitle {Near-Earth Sub-Alfv\'{e}nic Solar Winds: Interplanetary
  Origins and Geomagnetic Impacts} {Near-earth sub-alfv\'{e}nic solar winds:
  Interplanetary origins and geomagnetic impacts}.{\BBCQ}
\newblock
\APACjournalVolNumPages{The Astrophysical Journal}{926}{2}{135}.
\newblock
\begin{APACrefURL} \url{https://dx.doi.org/10.3847/1538-4357/ac4471}
  \end{APACrefURL}
\newblock
\begin{APACrefDOI} \doi{10.3847/1538-4357/ac4471} \end{APACrefDOI}
\PrintBackRefs{\CurrentBib}

\bibitem [\protect \citeauthoryear {%
Hasegawa%
, Sato%
\BCBL {}\ \BBA {} Akasofu%
}{%
Hasegawa%
\ \protect \BOthers {.}}{%
{\protect \APACyear {1979}}%
}]{%
Hasegawa1979}
\APACinsertmetastar {%
Hasegawa1979}%
\begin{APACrefauthors}%
Hasegawa, A.%
, Sato, T.%
\BCBL {}\ \BBA {} Akasofu, S.%
\end{APACrefauthors}%
\unskip\
\newblock
\APACrefYearMonthDay{1979}{}{}.
\newblock
{\BBOQ}\APACrefatitle {Dynamics of the Magnetosphere} {Dynamics of the
  magnetosphere}.{\BBCQ}
\newblock
\APACjournalVolNumPages{Edited by S.-l. Akasofu. Highham, Mass.: D.
  Reidel}{}{}{}.
\PrintBackRefs{\CurrentBib}

\bibitem [\protect \citeauthoryear {%
Iijima%
\ \BBA {} Potemra%
}{%
Iijima%
\ \BBA {} Potemra%
}{%
{\protect \APACyear {1976}}%
}]{%
Iijima1976}
\APACinsertmetastar {%
Iijima1976}%
\begin{APACrefauthors}%
Iijima, T.%
\BCBT {}\ \BBA {} Potemra, T\BPBI A.%
\end{APACrefauthors}%
\unskip\
\newblock
\APACrefYearMonthDay{1976}{}{}.
\newblock
{\BBOQ}\APACrefatitle {The amplitude distribution of field-aligned currents at
  northern high latitudes observed by Triad} {The amplitude distribution of
  field-aligned currents at northern high latitudes observed by triad}.{\BBCQ}
\newblock
\APACjournalVolNumPages{Journal of Geophysical Research
  (1896-1977)}{81}{13}{2165-2174}.
\newblock
\begin{APACrefDOI} \doi{https://doi.org/10.1029/JA081i013p02165}
  \end{APACrefDOI}
\PrintBackRefs{\CurrentBib}

\bibitem [\protect \citeauthoryear {%
Iijima%
\ \BBA {} Potemra%
}{%
Iijima%
\ \BBA {} Potemra%
}{%
{\protect \APACyear {1978}}%
}]{%
Iijima1978}
\APACinsertmetastar {%
Iijima1978}%
\begin{APACrefauthors}%
Iijima, T.%
\BCBT {}\ \BBA {} Potemra, T\BPBI A.%
\end{APACrefauthors}%
\unskip\
\newblock
\APACrefYearMonthDay{1978}{}{}.
\newblock
{\BBOQ}\APACrefatitle {Large-scale characteristics of field-aligned currents
  associated with substorms} {Large-scale characteristics of field-aligned
  currents associated with substorms}.{\BBCQ}
\newblock
\APACjournalVolNumPages{Journal of Geophysical Research: Space
  Physics}{83}{A2}{599-615}.
\newblock
\begin{APACrefDOI} \doi{https://doi.org/10.1029/JA083iA02p00599}
  \end{APACrefDOI}
\PrintBackRefs{\CurrentBib}

\bibitem [\protect \citeauthoryear {%
Kaeppler%
\ \protect \BOthers {.}}{%
Kaeppler%
\ \protect \BOthers {.}}{%
{\protect \APACyear {2015}}%
}]{%
Kaeppler2015}
\APACinsertmetastar {%
Kaeppler2015}%
\begin{APACrefauthors}%
Kaeppler, S\BPBI R.%
, Hampton, D\BPBI L.%
, Nicolls, M\BPBI J.%
, Strømme, A.%
, Solomon, S\BPBI C.%
, Hecht, J\BPBI H.%
\BCBL {}\ \BBA {} Conde, M\BPBI G.%
\end{APACrefauthors}%
\unskip\
\newblock
\APACrefYearMonthDay{2015}{}{}.
\newblock
{\BBOQ}\APACrefatitle {An investigation comparing ground-based techniques that
  quantify auroral electron flux and conductance} {An investigation comparing
  ground-based techniques that quantify auroral electron flux and
  conductance}.{\BBCQ}
\newblock
\APACjournalVolNumPages{Journal of Geophysical Research: Space
  Physics}{120}{10}{9038-9056}.
\newblock
\begin{APACrefDOI} \doi{https://doi.org/10.1002/2015JA021396} \end{APACrefDOI}
\PrintBackRefs{\CurrentBib}

\bibitem [\protect \citeauthoryear {%
Keiling%
, Thaller%
, Wygant%
\BCBL {}\ \BBA {} Dombeck%
}{%
Keiling%
\ \protect \BOthers {.}}{%
{\protect \APACyear {2019}}%
}]{%
Keiling2019}
\APACinsertmetastar {%
Keiling2019}%
\begin{APACrefauthors}%
Keiling, A.%
, Thaller, S.%
, Wygant, J.%
\BCBL {}\ \BBA {} Dombeck, J.%
\end{APACrefauthors}%
\unskip\
\newblock
\APACrefYearMonthDay{2019}{}{}.
\newblock
{\BBOQ}\APACrefatitle {Assessing the global Alfv\'{e}n wave power flow into and
  out of the auroral acceleration region during geomagnetic storms} {Assessing
  the global alfv\'{e}n wave power flow into and out of the auroral
  acceleration region during geomagnetic storms}.{\BBCQ}
\newblock
\APACjournalVolNumPages{Science Advances}{5}{6}{eaav8411}.
\newblock
\begin{APACrefDOI} \doi{10.1126/sciadv.aav8411} \end{APACrefDOI}
\PrintBackRefs{\CurrentBib}

\bibitem [\protect \citeauthoryear {%
Laundal%
\ \BBA {} Richmond%
}{%
Laundal%
\ \BBA {} Richmond%
}{%
{\protect \APACyear {2017}}%
}]{%
Laundal2017}
\APACinsertmetastar {%
Laundal2017}%
\begin{APACrefauthors}%
Laundal, K\BPBI M.%
\BCBT {}\ \BBA {} Richmond, A.%
\end{APACrefauthors}%
\unskip\
\newblock
\APACrefYearMonthDay{2017}{03}{}.
\newblock
{\BBOQ}\APACrefatitle {Magnetic Coordinate Systems} {Magnetic coordinate
  systems}.{\BBCQ}
\newblock
\APACjournalVolNumPages{Space Science Reviews}{206}{}{}.
\newblock
\begin{APACrefDOI} \doi{10.1007/s11214-016-0275-y} \end{APACrefDOI}
\PrintBackRefs{\CurrentBib}

\bibitem [\protect \citeauthoryear {%
Lavraud%
\ \protect \BOthers {.}}{%
Lavraud%
\ \protect \BOthers {.}}{%
{\protect \APACyear {2002}}%
}]{%
Lavraud2002}
\APACinsertmetastar {%
Lavraud2002}%
\begin{APACrefauthors}%
Lavraud, B.%
, Dunlop, M.%
, Phan, T.%
, R\`eme, H.%
, Bosqued, J.%
, Dandouras, I.%
\BDBL {}Balogh, A.%
\end{APACrefauthors}%
\unskip\
\newblock
\APACrefYearMonthDay{2002}{10}{}.
\newblock
{\BBOQ}\APACrefatitle {Cluster observations of the exterior cusp and its
  surrounding boundaries under northward IMF} {Cluster observations of the
  exterior cusp and its surrounding boundaries under northward imf}.{\BBCQ}
\newblock
\APACjournalVolNumPages{Geophysical Research Letters, v.29 (2002)}{29}{}{}.
\newblock
\begin{APACrefDOI} \doi{10.1029/2002GL015464} \end{APACrefDOI}
\PrintBackRefs{\CurrentBib}

\bibitem [\protect \citeauthoryear {%
Lin%
\ \protect \BOthers {.}}{%
Lin%
\ \protect \BOthers {.}}{%
{\protect \APACyear {2021}}%
}]{%
Lin2021}
\APACinsertmetastar {%
Lin2021}%
\begin{APACrefauthors}%
Lin, D.%
, Sorathia, K.%
, Wang, W.%
, Merkin, V.%
, Bao, S.%
, Pham, K.%
\BDBL {}Anderson, B.%
\end{APACrefauthors}%
\unskip\
\newblock
\APACrefYearMonthDay{2021}{}{}.
\newblock
{\BBOQ}\APACrefatitle {The Role of Diffuse Electron Precipitation in the
  Formation of Subauroral Polarization Streams} {The role of diffuse electron
  precipitation in the formation of subauroral polarization streams}.{\BBCQ}
\newblock
\APACjournalVolNumPages{Journal of Geophysical Research: Space
  Physics}{126}{12}{e2021JA029792}.
\newblock
\begin{APACrefDOI} \doi{https://doi.org/10.1029/2021JA029792} \end{APACrefDOI}
\PrintBackRefs{\CurrentBib}

\bibitem [\protect \citeauthoryear {%
Liu%
\ \protect \BOthers {.}}{%
Liu%
\ \protect \BOthers {.}}{%
{\protect \APACyear {2022}}%
}]{%
Liu2022}
\APACinsertmetastar {%
Liu2022}%
\begin{APACrefauthors}%
Liu, T\BPBI Z.%
, Wang, C\BHBI P.%
, Wang, X.%
, Angelopoulos, V.%
, Zhang, H.%
, Lu, X.%
\BCBL {}\ \BBA {} Lin, Y.%
\end{APACrefauthors}%
\unskip\
\newblock
\APACrefYearMonthDay{2022}{}{}.
\newblock
{\BBOQ}\APACrefatitle {Magnetospheric Field-Aligned Current Generation by
  Foreshock Transients: Contribution by Flow Vortices and Pressure Gradients}
  {Magnetospheric field-aligned current generation by foreshock transients:
  Contribution by flow vortices and pressure gradients}.{\BBCQ}
\newblock
\APACjournalVolNumPages{Journal of Geophysical Research: Space
  Physics}{127}{11}{e2022JA030700}.
\newblock
\begin{APACrefDOI} \doi{https://doi.org/10.1029/2022JA030700} \end{APACrefDOI}
\PrintBackRefs{\CurrentBib}

\bibitem [\protect \citeauthoryear {%
{Meitzler}%
\ \protect \BOthers {.}}{%
{Meitzler}%
\ \protect \BOthers {.}}{%
{\protect \APACyear {2023}}%
}]{%
Meitzler2023}
\APACinsertmetastar {%
Meitzler2023}%
\begin{APACrefauthors}%
{Meitzler}, R.%
, {Jun}, I.%
, {Blase}, R.%
, {Cassidy}, T.%
, {Clark}, R.%
, {Cochrane}, C.%
\BDBL {}{Yokley}, Z.%
\end{APACrefauthors}%
\unskip\
\newblock
\APACrefYearMonthDay{2023}{{\APACmonth{10}}}{}.
\newblock
{\BBOQ}\APACrefatitle {{Investigating Europa's Radiation Environment with the
  Europa Clipper Radiation Monitor}} {{Investigating Europa's Radiation
  Environment with the Europa Clipper Radiation Monitor}}.{\BBCQ}
\newblock
\APACjournalVolNumPages{Space Science Reviews}{219}{7}{61}.
\newblock
\begin{APACrefDOI} \doi{10.1007/s11214-023-01003-8} \end{APACrefDOI}
\PrintBackRefs{\CurrentBib}

\bibitem [\protect \citeauthoryear {%
Merkin%
\ \BBA {} Lyon%
}{%
Merkin%
\ \BBA {} Lyon%
}{%
{\protect \APACyear {2010}}%
}]{%
Merkin2010}
\APACinsertmetastar {%
Merkin2010}%
\begin{APACrefauthors}%
Merkin, V\BPBI G.%
\BCBT {}\ \BBA {} Lyon, J\BPBI G.%
\end{APACrefauthors}%
\unskip\
\newblock
\APACrefYearMonthDay{2010}{}{}.
\newblock
{\BBOQ}\APACrefatitle {Effects of the low-latitude ionospheric boundary
  condition on the global magnetosphere} {Effects of the low-latitude
  ionospheric boundary condition on the global magnetosphere}.{\BBCQ}
\newblock
\APACjournalVolNumPages{Journal of Geophysical Research: Space
  Physics}{115}{A10}{}.
\newblock
\begin{APACrefDOI} \doi{https://doi.org/10.1029/2010JA015461} \end{APACrefDOI}
\PrintBackRefs{\CurrentBib}

\bibitem [\protect \citeauthoryear {%
{Ni}%
\ \protect \BOthers {.}}{%
{Ni}%
\ \protect \BOthers {.}}{%
{\protect \APACyear {2016}}%
}]{%
Ni2016}
\APACinsertmetastar {%
Ni2016}%
\begin{APACrefauthors}%
{Ni}, B.%
, {Thorne}, R\BPBI M.%
, {Zhang}, X.%
, {Bortnik}, J.%
, {Pu}, Z.%
, {Xie}, L.%
\BDBL {}{Gu}, X.%
\end{APACrefauthors}%
\unskip\
\newblock
\APACrefYearMonthDay{2016}{{\APACmonth{04}}}{}.
\newblock
{\BBOQ}\APACrefatitle {{Origins of the Earth's Diffuse Auroral Precipitation}}
  {{Origins of the Earth's Diffuse Auroral Precipitation}}.{\BBCQ}
\newblock
\APACjournalVolNumPages{Space Science Reviews}{200}{1-4}{205-259}.
\newblock
\begin{APACrefDOI} \doi{10.1007/s11214-016-0234-7} \end{APACrefDOI}
\PrintBackRefs{\CurrentBib}

\bibitem [\protect \citeauthoryear {%
Papitashvili%
\ \BBA {} King%
}{%
Papitashvili%
\ \BBA {} King%
}{%
{\protect \APACyear {2020}}%
}]{%
omni1min}
\APACinsertmetastar {%
omni1min}%
\begin{APACrefauthors}%
Papitashvili, N\BPBI E.%
\BCBT {}\ \BBA {} King, J\BPBI H.%
\end{APACrefauthors}%
\unskip\
\newblock
\APACrefYearMonthDay{2020}{}{}.
\newblock
\APACrefbtitle {OMNI 1-min Data, NASA Space Physics Data Facility.} {Omni 1-min
  data, nasa space physics data facility.}
\newblock
\APACrefnote{Accessed: 2025-02-21}
\newblock
\begin{APACrefDOI} \doi{https://doi.org/10.48322/45bb-8792} \end{APACrefDOI}
\PrintBackRefs{\CurrentBib}

\bibitem [\protect \citeauthoryear {%
Pedersen%
\ \protect \BOthers {.}}{%
Pedersen%
\ \protect \BOthers {.}}{%
{\protect \APACyear {2021}}%
}]{%
Pederson2021}
\APACinsertmetastar {%
Pederson2021}%
\begin{APACrefauthors}%
Pedersen, M\BPBI N.%
, Vanham\"{a}ki, H.%
, Aikio, A\BPBI T.%
, K\"{a}ki, S.%
, Workayehu, A\BPBI B.%
, Waters, C\BPBI L.%
\BCBL {}\ \BBA {} Gjerloev, J\BPBI W.%
\end{APACrefauthors}%
\unskip\
\newblock
\APACrefYearMonthDay{2021}{}{}.
\newblock
{\BBOQ}\APACrefatitle {Field-Aligned and Ionospheric Currents by AMPERE and
  SuperMAG During HSS/SIR-Driven Storms} {Field-aligned and ionospheric
  currents by ampere and supermag during hss/sir-driven storms}.{\BBCQ}
\newblock
\APACjournalVolNumPages{Journal of Geophysical Research: Space
  Physics}{126}{11}{e2021JA029437}.
\newblock
\begin{APACrefDOI} \doi{https://doi.org/10.1029/2021JA029437} \end{APACrefDOI}
\PrintBackRefs{\CurrentBib}

\bibitem [\protect \citeauthoryear {%
Pilipenko%
}{%
Pilipenko%
}{%
{\protect \APACyear {2021}}%
}]{%
Pilipenko2021}
\APACinsertmetastar {%
Pilipenko2021}%
\begin{APACrefauthors}%
Pilipenko, V\BPBI A.%
\end{APACrefauthors}%
\unskip\
\newblock
\APACrefYearMonthDay{2021}{}{}.
\newblock
{\BBOQ}\APACrefatitle {Space weather impact on ground-based technological
  systems} {Space weather impact on ground-based technological systems}.{\BBCQ}
\newblock
\APACjournalVolNumPages{Solar-Terrestrial Physics}{7}{3}{68--104}.
\PrintBackRefs{\CurrentBib}

\bibitem [\protect \citeauthoryear {%
Ridley%
}{%
Ridley%
}{%
{\protect \APACyear {2007}}%
}]{%
Ridley2007}
\APACinsertmetastar {%
Ridley2007}%
\begin{APACrefauthors}%
Ridley, A\BPBI J.%
\end{APACrefauthors}%
\unskip\
\newblock
\APACrefYearMonthDay{2007}{}{}.
\newblock
{\BBOQ}\APACrefatitle {Alfv\'{e}n wings at Earth's magnetosphere under strong
  interplanetary magnetic fields} {Alfv\'{e}n wings at earth's magnetosphere
  under strong interplanetary magnetic fields}.{\BBCQ}
\newblock
\APACjournalVolNumPages{Annales Geophysicae}{25}{2}{533--542}.
\newblock
\begin{APACrefURL} \url{https://angeo.copernicus.org/articles/25/533/2007/}
  \end{APACrefURL}
\newblock
\begin{APACrefDOI} \doi{10.5194/angeo-25-533-2007} \end{APACrefDOI}
\PrintBackRefs{\CurrentBib}

\bibitem [\protect \citeauthoryear {%
Robinson%
, Vondrak%
, Miller%
, Dabbs%
\BCBL {}\ \BBA {} Hardy%
}{%
Robinson%
\ \protect \BOthers {.}}{%
{\protect \APACyear {1987}}%
}]{%
Robinson1987}
\APACinsertmetastar {%
Robinson1987}%
\begin{APACrefauthors}%
Robinson, R\BPBI M.%
, Vondrak, R\BPBI R.%
, Miller, K.%
, Dabbs, T.%
\BCBL {}\ \BBA {} Hardy, D.%
\end{APACrefauthors}%
\unskip\
\newblock
\APACrefYearMonthDay{1987}{}{}.
\newblock
{\BBOQ}\APACrefatitle {On calculating ionospheric conductances from the flux
  and energy of precipitating electrons} {On calculating ionospheric
  conductances from the flux and energy of precipitating electrons}.{\BBCQ}
\newblock
\APACjournalVolNumPages{Journal of Geophysical Research: Space
  Physics}{92}{A3}{2565-2569}.
\newblock
\begin{APACrefDOI} \doi{https://doi.org/10.1029/JA092iA03p02565}
  \end{APACrefDOI}
\PrintBackRefs{\CurrentBib}

\bibitem [\protect \citeauthoryear {%
Sato%
\ \BBA {} Iijima%
}{%
Sato%
\ \BBA {} Iijima%
}{%
{\protect \APACyear {1979}}%
}]{%
Sato1979}
\APACinsertmetastar {%
Sato1979}%
\begin{APACrefauthors}%
Sato, T.%
\BCBT {}\ \BBA {} Iijima, T.%
\end{APACrefauthors}%
\unskip\
\newblock
\APACrefYearMonthDay{1979}{}{}.
\newblock
{\BBOQ}\APACrefatitle {Primary sources of large-scale Birkeland currents}
  {Primary sources of large-scale birkeland currents}.{\BBCQ}
\newblock
\APACjournalVolNumPages{Space Science Reviews}{24}{3}{347--366}.
\PrintBackRefs{\CurrentBib}

\bibitem [\protect \citeauthoryear {%
Sorathia%
\ \protect \BOthers {.}}{%
Sorathia%
\ \protect \BOthers {.}}{%
{\protect \APACyear {2020}}%
}]{%
Sorathia2020}
\APACinsertmetastar {%
Sorathia2020}%
\begin{APACrefauthors}%
Sorathia, K\BPBI A.%
, Merkin, V\BPBI G.%
, Panov, E\BPBI V.%
, Zhang, B.%
, Lyon, J\BPBI G.%
, Garretson, J.%
\BDBL {}Wiltberger, M.%
\end{APACrefauthors}%
\unskip\
\newblock
\APACrefYearMonthDay{2020}{}{}.
\newblock
{\BBOQ}\APACrefatitle {Ballooning-Interchange Instability in the Near-Earth
  Plasma Sheet and Auroral Beads: Global Magnetospheric Modeling at the Limit
  of the MHD Approximation} {Ballooning-interchange instability in the
  near-earth plasma sheet and auroral beads: Global magnetospheric modeling at
  the limit of the mhd approximation}.{\BBCQ}
\newblock
\APACjournalVolNumPages{Geophysical Research Letters}{47}{14}{e2020GL088227}.
\newblock
\begin{APACrefDOI} \doi{https://doi.org/10.1029/2020GL088227} \end{APACrefDOI}
\PrintBackRefs{\CurrentBib}

\bibitem [\protect \citeauthoryear {%
Sorathia%
\ \protect \BOthers {.}}{%
Sorathia%
\ \protect \BOthers {.}}{%
{\protect \APACyear {2023}}%
}]{%
Sorathia2023}
\APACinsertmetastar {%
Sorathia2023}%
\begin{APACrefauthors}%
Sorathia, K\BPBI A.%
, Michael, A.%
, Merkin, V\BPBI G.%
, Ohtani, S.%
, Keesee, A\BPBI M.%
, Sciola, A.%
\BDBL {}Pulkkinen, A.%
\end{APACrefauthors}%
\unskip\
\newblock
\APACrefYearMonthDay{2023}{}{}.
\newblock
{\BBOQ}\APACrefatitle {Multiscale Magnetosphere-Ionosphere Coupling During
  Stormtime: A Case Study of the Dawnside Current Wedge} {Multiscale
  magnetosphere-ionosphere coupling during stormtime: A case study of the
  dawnside current wedge}.{\BBCQ}
\newblock
\APACjournalVolNumPages{Journal of Geophysical Research: Space
  Physics}{128}{11}{e2023JA031594}.
\newblock
\begin{APACrefDOI} \doi{https://doi.org/10.1029/2023JA031594} \end{APACrefDOI}
\PrintBackRefs{\CurrentBib}

\bibitem [\protect \citeauthoryear {%
Sorathia%
\ \protect \BOthers {.}}{%
Sorathia%
\ \protect \BOthers {.}}{%
{\protect \APACyear {2024}}%
}]{%
Sorathia2024}
\APACinsertmetastar {%
Sorathia2024}%
\begin{APACrefauthors}%
Sorathia, K\BPBI A.%
, Shumko, M.%
, Sciola, A.%
, Michael, A.%
, Merkin, V\BPBI G.%
, Gallardo-Lacourt, B.%
\BDBL {}Ukhorskiy, A\BPBI Y.%
\end{APACrefauthors}%
\unskip\
\newblock
\APACrefYearMonthDay{2024}{}{}.
\newblock
{\BBOQ}\APACrefatitle {Identifying the Magnetospheric Drivers of Giant
  Undulations: Global Modeling of the Evolving Inner Magnetosphere and Its
  Auroral Manifestations} {Identifying the magnetospheric drivers of giant
  undulations: Global modeling of the evolving inner magnetosphere and its
  auroral manifestations}.{\BBCQ}
\newblock
\APACjournalVolNumPages{Geophysical Research Letters}{51}{16}{e2024GL110772}.
\newblock
\begin{APACrefDOI} \doi{https://doi.org/10.1029/2024GL110772} \end{APACrefDOI}
\PrintBackRefs{\CurrentBib}

\bibitem [\protect \citeauthoryear {%
Toffoletto%
, Sazykin%
, Spiro%
\BCBL {}\ \BBA {} Wolf%
}{%
Toffoletto%
\ \protect \BOthers {.}}{%
{\protect \APACyear {2003}}%
}]{%
Toffoletto2003}
\APACinsertmetastar {%
Toffoletto2003}%
\begin{APACrefauthors}%
Toffoletto, F.%
, Sazykin, S.%
, Spiro, R.%
\BCBL {}\ \BBA {} Wolf, R.%
\end{APACrefauthors}%
\unskip\
\newblock
\APACrefYearMonthDay{2003}{}{}.
\newblock
{\BBOQ}\APACrefatitle {Inner magnetospheric modeling with the Rice Convection
  Model} {Inner magnetospheric modeling with the rice convection model}.{\BBCQ}
\newblock
\APACjournalVolNumPages{Space science reviews}{107}{}{175--196}.
\PrintBackRefs{\CurrentBib}

\bibitem [\protect \citeauthoryear {%
Waters%
\ \protect \BOthers {.}}{%
Waters%
\ \protect \BOthers {.}}{%
{\protect \APACyear {2020}}%
}]{%
Waters2020}
\APACinsertmetastar {%
Waters2020}%
\begin{APACrefauthors}%
Waters, C\BPBI L.%
, Anderson, B\BPBI J.%
, Green, D\BPBI L.%
, Korth, H.%
, Barnes, R\BPBI J.%
\BCBL {}\ \BBA {} Vanham{\"a}ki, H.%
\end{APACrefauthors}%
\unskip\
\newblock
\APACrefYearMonthDay{2020}{}{}.
\newblock
{\BBOQ}\APACrefatitle {Science Data Products for AMPERE} {Science data products
  for ampere}.{\BBCQ}
\newblock
\BIn{} M\BPBI W.~Dunlop\ \BBA {} H.~L{\"u}hr\ (\BEDS), \APACrefbtitle
  {Ionospheric Multi-Spacecraft Analysis Tools: Approaches for Deriving
  Ionospheric Parameters} {Ionospheric multi-spacecraft analysis tools:
  Approaches for deriving ionospheric parameters}\ (\BPGS\ 141--165).
\newblock
\APACaddressPublisher{Cham}{Springer International Publishing}.
\newblock
\begin{APACrefURL} \url{https://doi.org/10.1007/978-3-030-26732-2_7}
  \end{APACrefURL}
\newblock
\begin{APACrefDOI} \doi{10.1007/978-3-030-26732-2_7} \end{APACrefDOI}
\PrintBackRefs{\CurrentBib}

\bibitem [\protect \citeauthoryear {%
Wilder%
, Lopez%
, Eriksson%
, Pham%
\BCBL {}\ \BBA {} Lin%
}{%
Wilder%
\ \protect \BOthers {.}}{%
{\protect \APACyear {2019}}%
}]{%
Wilder2019}
\APACinsertmetastar {%
Wilder2019}%
\begin{APACrefauthors}%
Wilder, F\BPBI D.%
, Lopez, R\BPBI E.%
, Eriksson, S.%
, Pham, K.%
\BCBL {}\ \BBA {} Lin, D.%
\end{APACrefauthors}%
\unskip\
\newblock
\APACrefYearMonthDay{2019}{}{}.
\newblock
{\BBOQ}\APACrefatitle {The Relative Importance of Geoeffective Length Versus
  Alfv\'{e}n Wing Formation in the Saturation of the Ionospheric Reverse
  Convection Potential} {The relative importance of geoeffective length versus
  alfv\'{e}n wing formation in the saturation of the ionospheric reverse
  convection potential}.{\BBCQ}
\newblock
\APACjournalVolNumPages{Geophysical Research Letters}{46}{3}{1126-1131}.
\newblock
\begin{APACrefDOI} \doi{https://doi.org/10.1029/2018GL080639} \end{APACrefDOI}
\PrintBackRefs{\CurrentBib}

\bibitem [\protect \citeauthoryear {%
Zhang%
\ \protect \BOthers {.}}{%
Zhang%
\ \protect \BOthers {.}}{%
{\protect \APACyear {2019}}%
}]{%
Zhang2019}
\APACinsertmetastar {%
Zhang2019}%
\begin{APACrefauthors}%
Zhang, B.%
, Sorathia, K\BPBI A.%
, Lyon, J\BPBI G.%
, Merkin, V\BPBI G.%
, Garretson, J\BPBI S.%
\BCBL {}\ \BBA {} Wiltberger, M.%
\end{APACrefauthors}%
\unskip\
\newblock
\APACrefYearMonthDay{2019}{sep}{}.
\newblock
{\BBOQ}\APACrefatitle {{GAMERA}: A Three-dimensional Finite-volume {MHD} Solver
  for Non-orthogonal Curvilinear Geometries} {{GAMERA}: A three-dimensional
  finite-volume {MHD} solver for non-orthogonal curvilinear geometries}.{\BBCQ}
\newblock
\APACjournalVolNumPages{The Astrophysical Journal Supplement
  Series}{244}{1}{20}.
\newblock
\begin{APACrefURL} \url{https://doi.org/10.3847/1538-4365/ab3a4c}
  \end{APACrefURL}
\newblock
\begin{APACrefDOI} \doi{10.3847/1538-4365/ab3a4c} \end{APACrefDOI}
\PrintBackRefs{\CurrentBib}

\bibitem [\protect \citeauthoryear {%
Zmuda%
, Martin%
\BCBL {}\ \BBA {} Heuring%
}{%
Zmuda%
\ \protect \BOthers {.}}{%
{\protect \APACyear {1966}}%
}]{%
Zmuda1966}
\APACinsertmetastar {%
Zmuda1966}%
\begin{APACrefauthors}%
Zmuda, A\BPBI J.%
, Martin, J\BPBI H.%
\BCBL {}\ \BBA {} Heuring, F\BPBI T.%
\end{APACrefauthors}%
\unskip\
\newblock
\APACrefYearMonthDay{1966}{}{}.
\newblock
{\BBOQ}\APACrefatitle {Transverse magnetic disturbances at 1100 kilometers in
  the auroral region} {Transverse magnetic disturbances at 1100 kilometers in
  the auroral region}.{\BBCQ}
\newblock
\APACjournalVolNumPages{Journal of Geophysical Research
  (1896-1977)}{71}{21}{5033-5045}.
\newblock
\begin{APACrefDOI} \doi{https://doi.org/10.1029/JZ071i021p05033}
  \end{APACrefDOI}
\PrintBackRefs{\CurrentBib}

\end{thebibliography}

%
%
%
%
%

\end{document}